\documentclass[twocolumn,prl,showpacs,preprintnumbers,amsmath,amssymb]{revtex4}
\usepackage{graphicx}% Include figure files
\usepackage{dcolumn}% Align table columns on decimal point
\usepackage{bm}% bold math

\begin{document}

\title{Exchange shift of stripe domains 
in antiferromagnetically coupled multilayers
}
\author{N. S. Kiselev}
\thanks
{Corresponding author 
% IFW Dresden,
% Postfach 270116, D--01171 Dresden, Germany. 
% Tel.: +49-351-4659-542; Fax: +49-351-4659-537
}
\email{m.kyselov@ifw-dresden.de}
\author{I. E. Dragunov}
\author{U.K. R\"o\ss ler}
\email{u.roessler@ifw-dresden.de}
\author{ A.N.\ Bogdanov}

\affiliation{IFW Dresden, Postfach 270116, D-01171 Dresden, Germany}

\affiliation{Donetsk Institute for Physics and Technology, 
 83114 Donetsk, Ukraine}

\date{\today}

\begin{abstract}
{
Antiferromagnetically coupled multilayers
with perpendicular anisotropy, as 
[CoPt]/Ru, Co/Ir, Fe/Au, display ferromagnetic
stripe phases as the ground states.
It is theoretically shown
that the antiferromagnetic interlayer exchange
causes a relative shift of domains
in adjacent layers.
This ``exchange shift'' is
responsible for several recently observed effects:
an anomalous broadening of domain walls, the formation 
of so-called ``tiger-tail'' patterns,
and a ``mixed state'' of antiferromagnetic
and ferromagnetic domains
in [CoPt]/Ru multilayers.
The derived analitical relations between the values of the
shift and the strength of antiferromagnetic coupling provide an 
effective method for a quantitative determination
of the interlayer exchange interactions. 
}
\end{abstract}

%
% %%% PACS numbers

%\keywords{}%Use showkeys class option if keyword
                              %display desired
         
\maketitle

% \clearpage
%

Nanoscale superlattices of antiferromagnetically
coupled ferromagnetic layers have already become 
components of magnetoresistive devices.
They are considered as promising materials 
for the emerging spin electronics and high-density
storage technologies  \cite{Fullerton03}.
An interesting group of these \textit{artificial} antiferromagnets
belongs to systems with high perpendicular anisotropy
(e. g. Co/Ru, Co/Ir, [Co/Pt]Ru, [Co/Pt]NiO
superlattices)
\cite{Hamada02,Hellwig03,Itoh03,Baruth06}.
Due to the strong competition between
antiferromagnetic interlayer exchange and 
magnetostatic couplings \cite{Hellwig03,stripes},
nanoscale superlattices with strong perpendicular
anisotropy display specific multidomain states and unusual 
magnetization processes
\cite{Hamada02,Hellwig03,Baruth06,Hellwig07},
which have no counterpart in other layered systems
with perpendicular magnetization \cite{Hubert98}.

So far, theoretical analysis of 
magnetization states and processes in 
antiferromagnetically coupled multilayers
with out-of-plane magnetization 
has been based on micromagnetic 
models of stripe domains,
where the domain walls 
throughout the whole stack of the ferromagnetic layers 
sit exactly on top of each other 
\cite{Hellwig03,stripes}.
In our letter we show that this 
assumption is wrong.
The antiferromagnetic interlayer coupling causes
a lateral shift of the domain walls 
in the adjacent ferromagnetic layers.
We develop a phenomenological theory of 
these complex stripe states.
The analytical evaluation of 
a basic two-layer model shows that 
the formation and evolution 
of such ``shifted'' multidomain phases 
should appreciably influence the appearance 
and the magnetization processes of stripe states 
in perpendicular, antiferromagnetically coupled multilayers.
%
%This ``exchange shift'' explains a number of effects 
%recently observed in these systems.

\begin{figure}
\includegraphics[width=8.5cm]{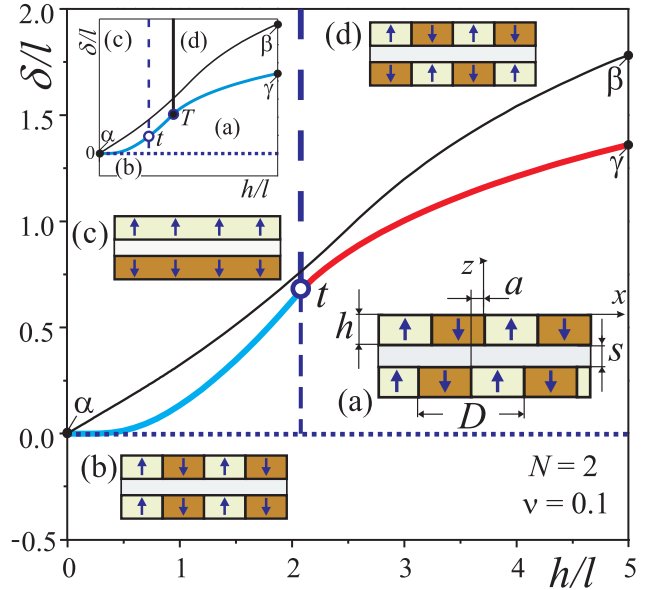}
\caption{
\label{PhD}
(Color online)
The magnetic phase diagram of
states in reduced variables for layer thickness
$h/l$ and interlayer exchange $\delta/l$ 
for $N = 2$ and $\nu = s/h =0.1$.
Thick lines indicate the first-order transitions
from the \textit{shifted} ferro stripe phase (a) into
the antiferro stripe (b) ($t-\gamma$) 
and homogeneous (c) phases
($\alpha-t$). These critical lines 
meet in a special critical point $t$
($h_t = 2.0816 l$, $\delta_t = 0.6385 l$), 
where a continuous transition line 
ends on a first-order line 
(hollow circle).
The thin solid line  $\alpha-\beta$ 
indicates the stability limit of 
the shifted ferro stripe phase. 
The dotted line ($\delta= 0$) 
indicates the second-order
transition from the shifted stripes ($a > 0$) 
in antiferromagnetically
coupled systems ($\delta>0$) into
the  ferro stripes (b) with  $ a = 0$ 
in ferromagnetically coupled or decoupled
multilayers ($\delta \le 0$).
At the dashed line $h \equiv h_t$ 
the antiferro phase (d)
continuously 
transforms into the homogeneous
antiferromagnetic states (c)
for $\delta > \delta_t$. 
The dashed line for $\delta < \delta_t$
is the stability limit of 
the metastable antiferro stripe phase.
Inset: sketch of the generic phase diagram
for $N\ge 4$ with the 
discontinuous transition of antiferro
stripes into the homogeneous state.
Point $T$ is a triple point.
}
\end{figure}

As a model we consider stripe domains in a superlattice
composed of $N$ identical layers of thickness
$h$ \textit{antiferromagnetically} coupled
via a spacer of thickness $s$.
The stripe domain phase consists of domains with alternate
magnetization $\mathbf{M}$ along the $z$-axis perpendicular
to the multilayer plane. The domains 
are separated by thin domain walls with
a finite area energy density $\sigma$.
The magnetic energy density of the model
(Fig. \ref{PhD} (a) ) can be written as
a function of the stripe period 
$D$ and the shift $a$
\begin{eqnarray}
W  = 
 \frac{2 \sigma N}{D} +2 \pi M^2 N \, w_m 
 \pm \frac{J}{h} \left(1- \frac{4 a}{D}\right)
 (1-N) 
 \,.
\label{energy0}
\end{eqnarray}
The first term in (\ref{energy0}) describes
the domain wall energy, $w_m$ is the stray field
energy, $J > 0$ is the antiferromagnetic 
exchange interaction.
The upper (lower) sign corresponds 
to an (anti)parallel arrangement of
the magnetization in the adjacent layers.
We call these modes \textit{ferro} 
and  \textit{antiferro} stripe phases.

We introduce a set of
reduced geometrical parameters 
\begin{eqnarray}
\label{pu}
p = 2\pi h/D,  \quad
u = 2a/D, \quad
\nu  = s/h
\end{eqnarray}
and two characteristic lengths
\begin{eqnarray}
l = \sigma /(4 \pi M^2), \quad
\delta = J/ (2 \pi M^2)
\label{l}
\end{eqnarray}
describing the relative energy contributions
of the domain walls
($l$) and the interlayer coupling ($\delta$)
in comparison to the stray field energy.
Then, the reduced energy $w=W/(2\pi M^2 N)$
can be written 
\begin{eqnarray}
\label{energy1}
w (p,u)  =4 p \frac{l}{h}
+\frac{\delta}{h}\left(1 -\frac{1}{N} \right)
\left( 1 - 2 u \right)+ w_m (p, u) \,.
\end{eqnarray}
The stray field energy $w_m (p, u)$
is derived by solving
the corresponding magnetostatic problem
\begin{eqnarray}
w_m  = \frac{8}{\pi ^2 p} 
\sum_{{\rm{odd}}\,n }^{\infty}  \frac{1}{n^3}
 \left[ \left( 1 - e^{ - np}  \right) - 
\sum_{k=1}^{N-1} f_k^{(n)}(p,u_k)\right]\,,
\label{stray1}
\end{eqnarray}
where $f_k^{(n)}(p,u_k)= 
2\cos \left( {\pi n \,u_k } \right) 
\sinh^2 \left( n p/2 \right) \exp ( -\tau np)$,
$\, u_k = u [1-(-1)^{k}]/2$, and 
$\, \tau = 1 + \nu$.

The identity 
$ \int_0^{\infty} t^{(m-1)}\exp(-nt) dt = (m-1)!/n^m$
allows one to transform
the infinite sums in Eq.~(\ref{stray1})
into integrals on the interval
$[0,1]$ (for details, see similar calculations
in \cite{FTT80,stripes})
\begin{eqnarray}
w_m  & = &  1 + \frac{4p}{\pi ^2 }\int\limits_0^1 
{\left( {1 - t} \right)} \ln \left[ {\tanh  
\left( \frac{pt}{2} \right)} \right]dt
\nonumber \\
& & +\sum_{k=1}^{N-1}
\left( 1 - \frac{k}{N} \right)\Xi_k \,(p,u_k) 
\,,
\label{stray2a}
\end{eqnarray}
where
$\Xi_k (p,u_k) = 2\Omega(p, u_k, \tau k) 
-\Omega (p, u_k,\tau k+1) -\Omega_{\nu+1} 
(p, u_k,\tau k -1 )$, and
\begin{eqnarray}
\Omega(p,u_k, \omega) = 
4 \left(\frac{\omega^2p}{\pi^2}- \frac{u_k^2}{p} \right)
I_{\omega}^{(1)}+
\frac{8 \omega u_k}{\pi}
I_{\omega}^{(2)}
\,,
\label{stray3}
\end{eqnarray}
\begin{eqnarray}
I_{\omega}^{(1)} (p, u_k) =
\int_0^1 (1-t)\,{\rm{arctanh}} \left[\frac{\cos(\pi u_k t)}
{\cosh(\omega p t)}\right] dt
\,,
\label{stray4}
\end{eqnarray}
\begin{eqnarray}
I_{\omega}^{(2)} (p, u_k) =
\int_0^1 (1-t)\,\arctan \left[\frac{\sin(\pi u_k t)}
{\sinh(\omega p t)}\right] dt.
\label{stray5}
\end{eqnarray}
%
%The reduced energy $w$ (\ref{energy1}) 
%includes two  internal variables ($p$,$u$)
%and three controle parameters
%($h/l$, $\delta/l$, $\nu$).
%
Minimization of $w$  with respect to
$p$ and $u$ yields
the equilibrium geometrical 
parameters, $D$ and  $a$, for the stripe domains 
as functions of
the four control parameters $h/l$, 
$\delta/l$, $\nu$, and $N$ in the model.
The phase diagram in variables ($h/l$, $\delta/l$) for $\nu = 0.1$
and $N = 2$ plotted in Fig. \ref{PhD} demonstrates
the main features of these solutions.
Depending on the values of
the materials parameters 
one of the four ground states
is realized in the system:
the ferro (b) or antiferro (d) stripes,
the shifted ferro stripe phase (a), or
the antiferromagnetic single domain state
(c).

\begin{figure}
\includegraphics[width=8.5cm]{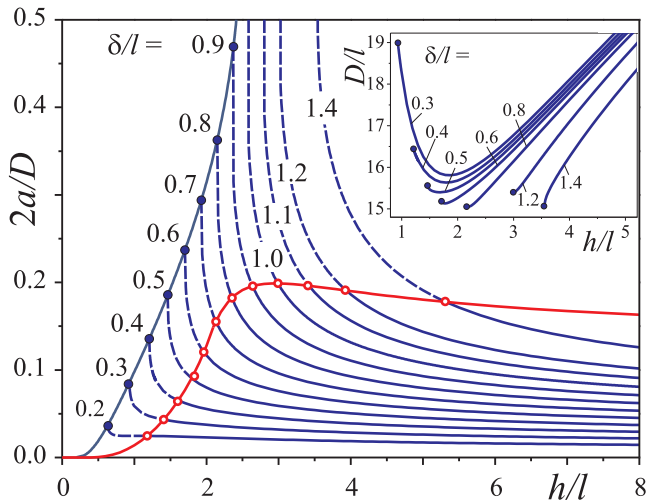}
\caption{
\label{solutions}
(Color online)
The equilibrium  values
of the reduced shift $2a/D$ 
and reduced period (inset) as functions
of the reduced thickness $h/l$ for
different values of $\delta = J/(2 \pi M^2)$
in a two-layer ($N=2$) with $\nu = 0.1$.
Hollow points indicate the solutions
at the transition line $\alpha-t-\gamma$, 
and solid points show the shift at the
lability line $\alpha-\beta$.
}
\end{figure}

The analytical results presented in Figs.~\ref{PhD}
and \ref{solutions} exemplify a fundamental
difference between multidomain states
in antiferromagnetically coupled 
superlattices ($\delta > 0$) 
and those in multilayers
with a ferromagnetic interlayer exchange
($\delta < 0$)
or in decoupled nanolayers 
($\delta = 0$).
In the latter cases the ferro stripe phase 
is the ground state for arbitrary
values of the control parameters
$\delta, h, \nu, N$ \cite{stripes}.
In the antiferromagnetic case the ferro stripes
can exist as stable or metastable state 
(i) only in a certain range of the control 
parameters (in Fig.~\ref{PhD}
below lability line $\alpha-\beta$), and (ii) 
the ferro stripes are unstable with respect 
to lateral shifts of domain walls in the
adjacent layers. 
%
%Thus, in antiferromagnetically
%coupled superlattices stripe domains
%with parallel arrangement in the adjacent layers
%only exist as \textit{shifted} ferro stripes.

%
At the critical line $\alpha-t-\gamma$ this
phase transforms into the homogeneous
(with $h < h_{t} = 2.0816\, l$ for $N=2$) 
or antiferro stripe phase ($h > h_{t}$)
by a first-order transition.
The ferro stripe mode is the
ground state of the system
between the transition lines $\alpha-t-\gamma$
and $ \delta =0$.
In the case $N=2$, 
the antiferro stripe phase 
transforms into the homogeneous
phase by the unlimited expansion
of the stripe period at the critical line $h \equiv h_{t}$ 
down to the critical point at $\delta_t=0.6385\,l$.
Below the transition line $t-\gamma$
the antiferro stripes (with zero shift)
still exist as metastable
states down to the line $\delta \equiv 0$.
In antiferromagnetic systems with $N\ge 4$,
the transition between the antiferromagnetic monodomain
and the antiferro stripe phase becomes discontinuous \cite{stripes}.
The general topology of these phase diagrams
(inset Fig.~\ref{PhD}) display triple points $T$,
where all three ground states co-exist, and the
antiferro stripe phase is only metastable at the 
special critical point $t$.
The ``exchange shift'' appreciably influences 
the appearance of the ferro stripe phase.
As shown in Fig.~\ref{solutions},
the shift $a$ can attain sizeable values.
%
%In observations on multidomain patterns, 
%this effect will be visible as  a ``broadening
%of the domain walls''.
%
Note that in antiferromagnetically
coupled superlattices the exchange energy 
of the shifted ferro stripes includes 
a negative contribution linear with respect
to $a$ (\ref{energy0}).
This is the mathematical reason for the instability
of the solutions with zero shift.
To elucidate this phenomenon we consider
small shift distortions in an isolated domain wall.
The perturbation energy (per domain wall length) 
$\Delta E (a) = \pm 2 \pi M^2 e(a)$
can be written as a series with respect to
the small parameter $a \ll s$ 
(the upper (lower) sign corresponds 
to (anti)ferro stripe states)
\begin{eqnarray}
e(a) =- 4 \delta a+ A(\nu)  a^2 
- B (\nu) s^{-2}  a^4
\,,
\label{Eshift}
\end{eqnarray}
where
$A(\nu)=(2/\pi) 
\ln \left[ (\nu+1)^2/(\nu (\nu+2)) \right]$,
$B (\nu) = (2+ 6 \nu + 3 \nu^2)/[3 \pi (\nu+1)^2 (\nu +2)^2]$.
The interlayer exchange coupling
energy in Eq.~(\ref{Eshift}) is linear
with respect to the shift $a$
and  \textit{negative} in the case
of the ferro stripes. 
This energy contribution
yields solutions with finite 
$a = 2 \delta A^{-1}(\nu)$ 
for arbitrary strengths of
the antiferromagnetic exchange.
On the contrary, for the antiferro stripes
this energy is \textit{positive} and 
the solution with \textit{zero} shift remains stable.
\begin{figure}
\includegraphics[width=8.5cm]{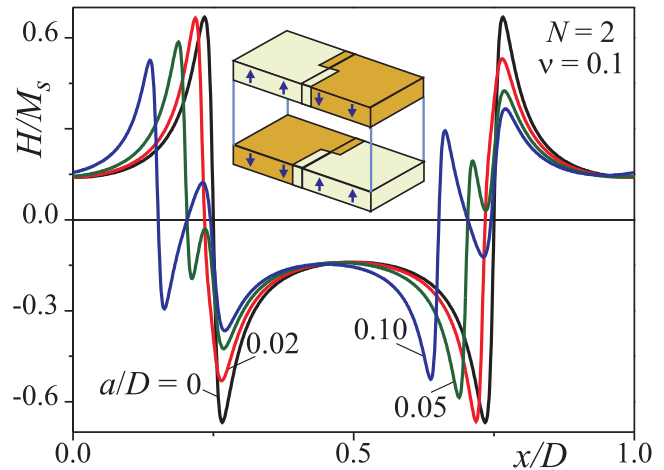}
\caption{
(Color online) 
The calculated stray-field profiles 
$H_z^{(m)}(x/D)$ for shifted ferro stripe
modes for a multilayer with $N = 2$ and
$\nu = 0.1$.
Inset: an isolated antiferromagnetic domain wall 
with ``tiger-tail'' distortion, the pattern is 
periodically repeated along the wall.
Such patterns act as nucleation lines
for ferro stripe modes.
\label{profiles}
}
\end{figure}
The analysis shows that for $h > h_t$
the antiferromagnetic stripes  
exist as metastable state between the transition 
lines $t-\gamma$ and  $\delta =0$.

In Ref.~\cite{Hellwig03}, experimental
domain observations are reported 
on antiferromagnetic [CoPt]/Ru multilayers 
with $N = 2$ to 10,
the magnetization $M = 700$ emu, 
interlayer exchange $ J = 0.45 $ erg/cm$^2$, 
and the parameter $\nu$ in the a range from 0.072 to 0.6. 
For these systems, the values of the shift $a$ 
vary from  $a = 4.5$ nm ($\nu = 0.72$) 
to $a = 18.5$ nm ($\nu = 0.6$).
These shifts amount to noticeable parts of the domain size,
$D/2 \simeq 130$ nm \cite{Hellwig03}).
The mathematical connection between values of 
the shift $a$ and the exchange constant $J$
can be used for an experimental determination
of the strength of the antiferromagnetic coupling.

The stray-field distributions above the superlattice
surfaces are highly sensitive to the appearance of
sizeable shifts $a$. Such effects can be investigated
by magnetic force microscopy.
Following  \cite{MFM} one can derive an analytical solution
for the stray fields  $\mathbf{H}^{(m)}$
above an antiferromagnetic two-layer 
in the shifted ferro stripe states.
E.g., the longitudinal
component $H_z^{(m)}$ is
\begin{eqnarray}
H_z^{(m)} & = &4 M [ \Upsilon (x, z)+ \Upsilon (x + a, z+h+s)
\nonumber \\
&-& \Upsilon (x, z + h) - \Upsilon (x+a, z +2h + s) ]
\,,
\label{strayfieldz}
\end{eqnarray}
where $\Upsilon (x, z) = \arctan[\cos(2 \pi x/D)/\sinh(2 \pi z//D)]$.
Peculiarities of the stray-field profiles $H_z^{(m)}(x/D)$ 
imposed by the exchange shift, as seen in Fig.~\ref{profiles},
should be measurable by magnetic force microscopy imaging.

For $h < h_t$ metastable isolated
domain walls can exist within 
the antiferromagnetically coupled 
ground state (Fig.~\ref{PhD}(c)).
Usually antiferromagnetic domain patterns 
are formed during demagnetization cycles 
in multilayers with the single domain antiferromagnetic
ground state (phase (c) in Fig.~\ref{PhD}),
see, e.g., 
Refs.~\cite{Hamada02,Hellwig03,
Baruth06,Hellwig07}.
These isolated walls correspond 
to the solutions with zero shift 
and preserve their (local)
stability down 
to vanishing antiferromagnetic coupling $\delta =0$.
The results of this paper elucidate the 
nature of so-called ``tiger-tail'' patterns
visible along these isolated domain walls 
of the antiferromagnetic phase
\cite{Baruth06,Hellwig07}
and recently observed as a ``mixed state''
of antiferro and ferro stripes \cite{Hellwig07}.
Isolated domain walls within the homogeneous
antiferromagnetic phase can play 
the role of nucleation
centers for the ferro stripe phase.
Within the metastability 
region of the shifted ferro stripe phase
(area between $\alpha-\beta$ and 
$\alpha-\gamma$ lines in Fig.~\ref{PhD})
sinusoidal distortions of antiferromagnetic domain walls
transform into spin configurations corresponding 
to the ferro stripe phase (Inset in Fig.~\ref{profiles}). 
Such patterns have been reported
in Ref.~\cite{Hellwig07} and were called ``tiger-tails''.
During the first-order phase transition 
at the $\alpha-\gamma$  line of the phase diagram, where
the monodomain antiferomagnetic phase and the shifted 
ferro stripes coexist, ``tiger-tails''
develop into ferro stripe patterns.
The transformation  of ``tiger-tails''
into extended areas with the ferro stripe phase
was observed in Ref.~\cite{Hellwig07}.
This is a ``mixed state'' composed of 
the homogeneous antiferromagnetic 
and the multidomain ferromagnetic phases. 
This particular evolution at a first-order transition,
when magnetic phases nucleate within domain walls of
competing phases have previously been observed 
in various bulk magnetic systems 
\cite{UFN88}.

In conclusion, we have demonstrated that,
in antiferromagnetically coupled multilayers,
ferromagnetic stripes are unstable with respect
to a lateral shift. 
These multidomain configurations 
form \textit{shifted ferro stripe} states
(Fig.~\ref{PhD} (a)). 
On the contrary,
the inhomogeneous magnetic states 
with antiparallel arrangement
in adjacent layers (antiferro stripes (d)
and isolated antiferromagnetic domain walls)
have stable configurations with zero shift \cite{Baruth2}.
 
\begin{acknowledgments}
Work supported by DFG through SPP1239 (project A08).
N.S.K., I.E.D., A.N.B.
\ thank H.\ Eschrig for support and
hospitality at IFW Dresden. 
\end{acknowledgments}


\begin{thebibliography} {99}
\bibitem{Baruth2}
In Ref. \cite{Baruth06} the solutions
for shifted antiferromagnetic domain walls
are claimed to exist.
%
However, analysis shows that the  solutions 
obtained in Ref. \cite{Baruth06} are related to
the shifted ferro stripe phase.
\end{thebibliography}
\end{document}